\documentclass{moriond}

\usepackage{amsmath}
\usepackage{xspace}

\def\be{\begin{equation}}
\def\ee{\end{equation}}
\def\bea{\begin{eqnarray}}
\def\eea{\end{eqnarray}}

\def\ROO{\ensuremath{R_\text{OO}}\xspace}
\def\RpO{\ensuremath{R_\text{pO}}\xspace}
\def\ROOpOpO{\ensuremath{R_\text{OO}/R_\text{pO}^2}\xspace}
\def\pt{\ensuremath{p_\text{T}}\xspace}
\def\piz{\ensuremath{\pi^\text{0}}\xspace}
\newcommand{\fivethreesixnn}{$\sqrt{s_{\mathrm{NN}}}~=~5.36$~Te\kern-.1emV\xspace}

\newcommand{\Photo}{\includegraphics[height=35mm]{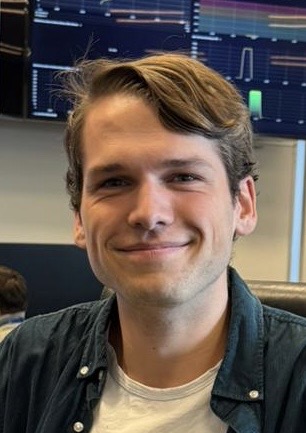}}

\begin{document}
\vspace*{4cm}
\title{Nuclear Modification of $\pi^0$ Production in OO Collisions with ALICE}

\author{ Nicolas Strangmann -- on behalf of the ALICE collaboration }

\address{Institute for Nuclear Physics, University of Frankfurt\\60438 Frankfurt am Main, Germany}

\maketitle\abstracts{
We present the first results on the $\pi^0$ nuclear modification factor \ROO in OO collisions at LHC energies by the ALICE experiment. The measurement of the modification of hadron production in nuclear collisions compared to a vacuum baseline in pp collisions is a valuable probe for parton energy loss in the hot medium. The ALICE \ROO results show significant (up to 4$\sigma$) suppression of $\pi^0$ production in OO collisions compared to the pp reference, and up to 2.4$\sigma$ deviation w.r.t. model predictions that include only cold nuclear matter eﬀects.}

\section{Introduction}

Ultra-relativistic collisions of heavy ions can lead to the formation of a state of deconfined quarks and gluons, known as the quark-gluon plasma (QGP). 
As this medium is expected to expand, cool and re-hadronize on very short timescales ($\mathcal{O}(10^{-23}\,\mathrm{s})$)~\cite{ALICE:2022wpn}, its properties are inferred from final-state observables.
Several signatures traditionally attributed to the presence of a QGP, such as strangeness enhancement and elliptic flow, have also been observed in high multiplicity pp and p--Pb collisions.
These findings challenge the assumption of small systems as a QGP-free reference for heavy-ion collisions. However, one key signature of the QGP has so far not been established outside of heavy-ion collisions: parton energy loss.
The energy lost by hard-scattered partons traversing the medium, and subsequently transferred to final-state hadrons through hadronization, can be quantified via the nuclear modification factor
\bea
R_\text{AA} = \frac{Y_\text{AA}}{\langle T_\text{AA} \rangle \sigma_\text{pp}},
\label{eq:RAA}
\eea
where $Y_\text{AA}$ denotes the particle yield in nucleus-nucleus collisions, $\sigma_\text{pp}$ the production cross section in pp collisions, and $\langle T_\text{AA} \rangle$ the average nuclear overlap function obtained from Glauber model calculations.
Measurements of $R_\text{AA}$ in heavy-ion collisions exhibit a strong suppression relative to unity, while no such suppression has been observed in p--Pb collisions. A dedicated oxygen-oxygen (OO) run at the LHC in July 2025 provides a unique opportunity to bridge the gap between small and large systems and probe the onset of QGP formation in an intermediate-sized system.
In this contribution, the measurement of $R_\text{OO}$ using neutral pions is presented and compared to theoretical predictions with and without parton energy loss.

\section{Detector setup and signal extraction}

The presented analysis utilizes two of ALICE's subdetectors. The Fast Interaction Trigger (FIT) system provides the trigger input for coincident signals in its two Cherenkov arrays located on either side of the interaction point. 
The Electromagnetic Calorimeter (EMCal) measures electromagnetic clusters, for which a minimum energy of 600\,MeV is required in this analysis. Photon-induced clusters are selected based on their shower shape and combined into neutral pion candidates by pairing photon candidates within the same event. The combinatorial background, estimated using a rotation method, is subtracted from the invariant mass distribution, and the raw signal is extracted by integrating around the $\pi^0$ mass to obtain the raw yield $N_{\pi^0}$.
The invariant production cross section in intervals of transverse momentum $\Delta p_\mathrm{T}$ and rapidity $\Delta y$ is obtained via
\bea
E \frac{\text{d}^3 \sigma_{\piz}}{\text{d}p^3} &= \frac{N_{\piz}}{N_\text{evt}}(1-S) \frac{1}{A \varepsilon_\text{rec} 2\pi \Delta y}\frac{1}{\pt\Delta\pt}\frac{1}{B}\times \sigma_\text{vis},
\label{eq:XSec}
\eea
where $N_\text{evt}$ denotes the number of recorded collisions (1.4B in pp, 2.1B in OO). The secondary $\pi^0$ contribution ($\sim4\%$) from long-lived decays is subtracted to obtain the primary yield. 
The yield is furthermore corrected for the geometric acceptance ($A$), reconstruction efficiency ($\varepsilon_\text{rec}$) and branching ratio $B$.
In pp collisions, the production cross section is obtained using the visible cross section $\sigma_\text{vis}$, while in OO collisions the invariant yield is corrected using the trigger efficiency determined from simulations.

\section{Results}

\subsection{Production of neutral pions in pp and OO collisions}

\begin{figure}
\centerline{\includegraphics[width=0.45\linewidth]{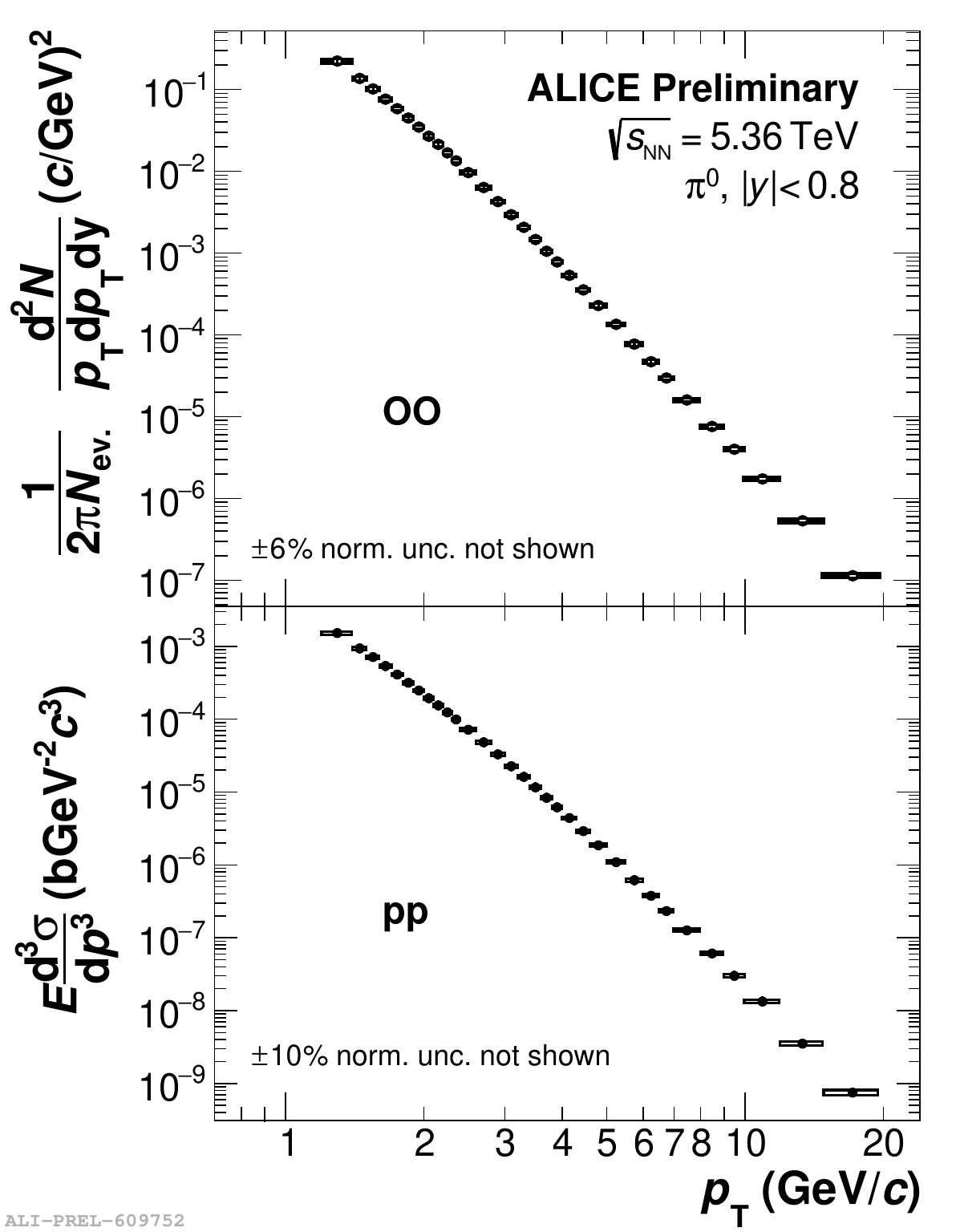}}
\caption[]{Measured invariant yield and cross section of neutral pions in inelastic OO and pp collisions at \fivethreesixnn within $\vert y \vert < 0.8$.}
\label{fig:Spectra}
\end{figure}

Figure \ref{fig:Spectra} shows the measured invariant yield of neutral pions in inelastic OO collisions (upper panel) and the production cross section in inelastic pp collisions (lower panel). 
The spectra are extracted within $\vert y \vert < 0.8$ and $1.2 < \pt$ (GeV/$c)<20$. Despite the short run duration of less than a week, the statistical uncertainties are below 1\% for a majority of the data points. Systematic uncertainties amount to approximately 5\%, dominated by the material budget uncertainty (4.2\%), which, however, cancels completely on the extracted \ROO. A conservative normalization uncertainty of 6\% (10\%) in OO (pp) is assumed while the Van der Meer scan analysis is ongoing.

\subsection{Measured nuclear modification factor}

\begin{figure}
\begin{minipage}{0.49\linewidth}
\centerline{\includegraphics[width=\linewidth]{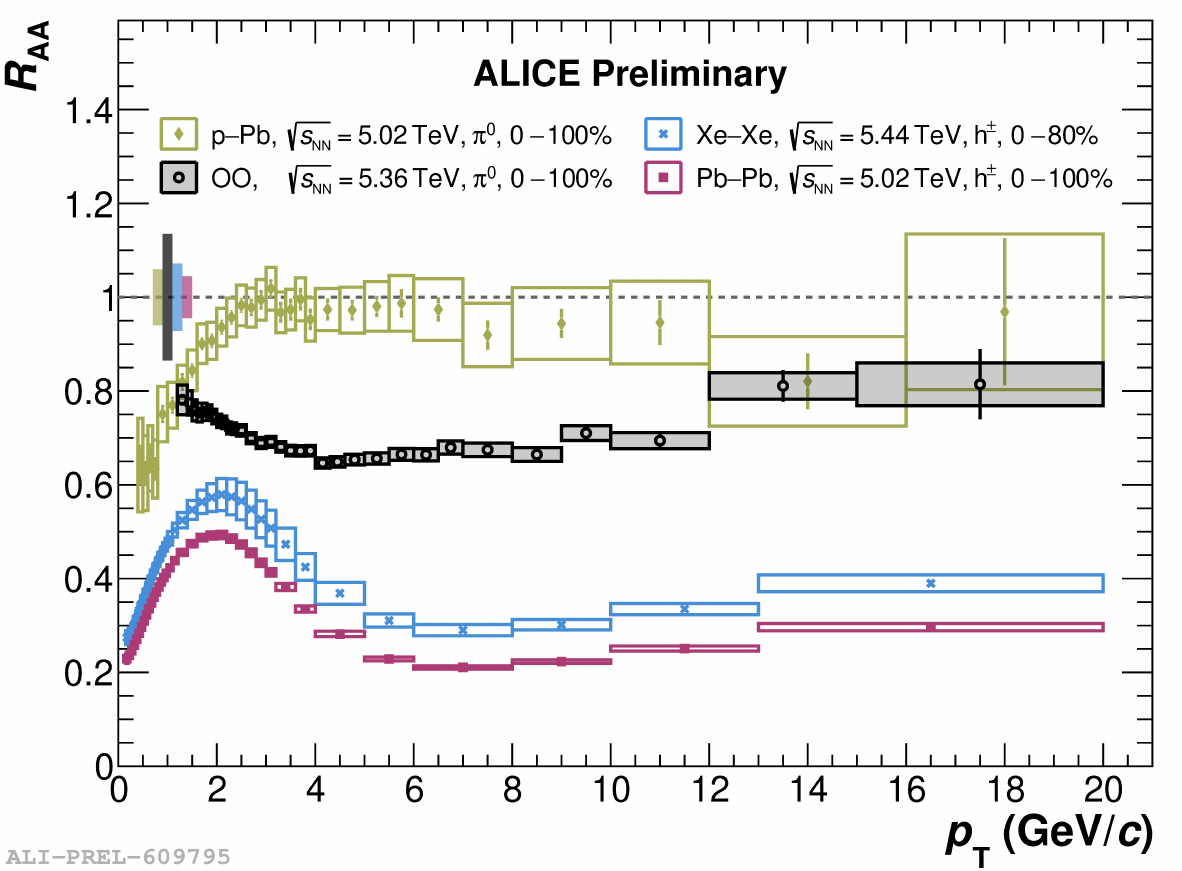}}
\end{minipage}
\hfill
\begin{minipage}{0.49\linewidth}
\centerline{\includegraphics[width=\linewidth]{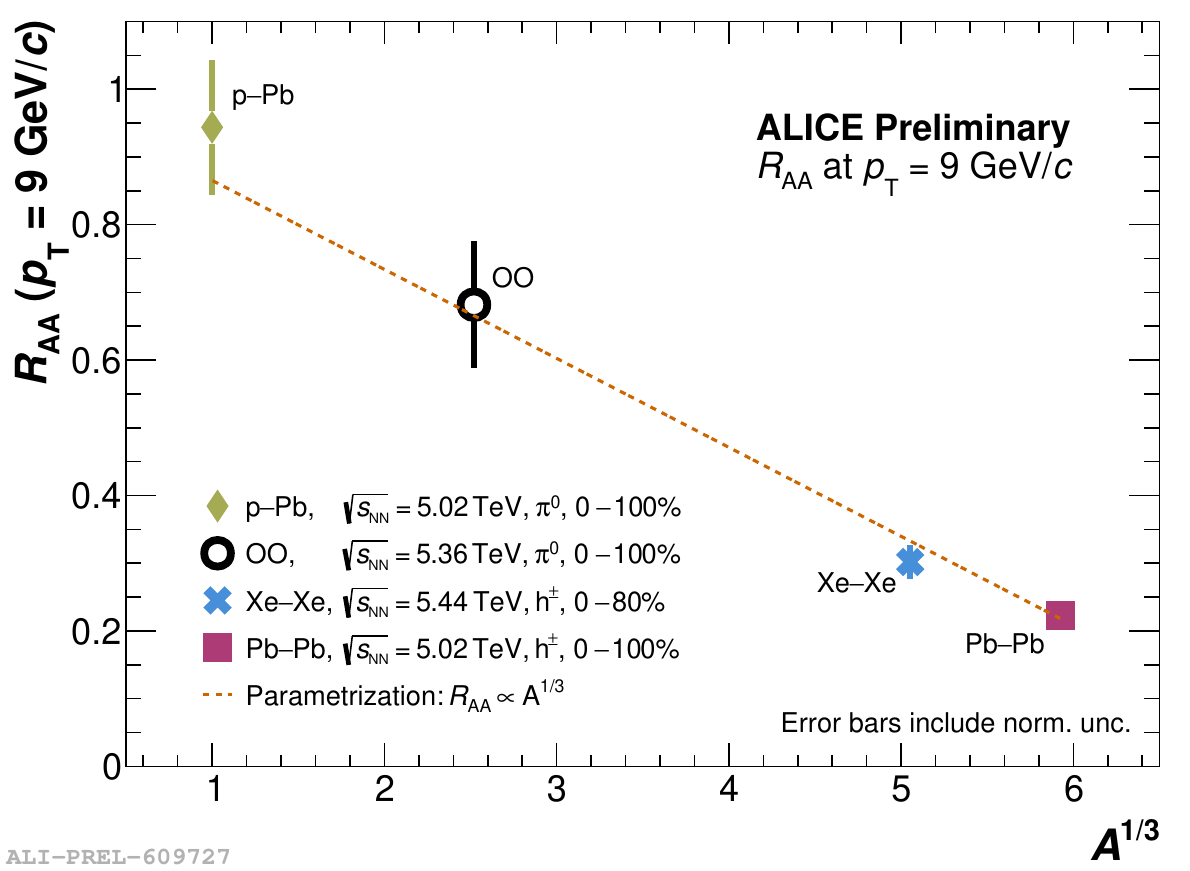}}
\end{minipage}
\caption[]{Nuclear modification factor $R_\text{OO}$ (black markers) together with the $R_\text{AA}$ in Pb--Pb~\cite{ALICE:2018ekf}, Xe--Xe~\cite{ALICE:2018hza} and p--Pb~\cite{ALICE:2018vhm} collisions. The left figure shows the \pt dependence of the modification, the right figure shows the value of $R_\text{AA}$ at \pt = 9 GeV/$c$ against the system size, quantified via $A^{1/3}$. }
\label{fig:ROOMeas}
\end{figure}

The spectra shown in Fig.~\ref{fig:Spectra} are compared according to Eq.~\ref{eq:RAA} to obtain the nuclear modification factor $R_\text{OO}$, shown in Fig.~\ref{fig:ROOMeas} (black markers).
The measured \ROO exhibits a suppression relative to unity with a significance of 4$\sigma$, and shows a \pt-shape similar to that observed in Pb--Pb and Xe--Xe collisions.
In particular, the characteristic low-\pt enhancement and high-$p_\mathrm{T}$ rise resembles the behavior seen in larger systems, while differing from the trend observed in p--Pb collisions. This suggests that cold nuclear matter effects alone are unlikely to explain the observed suppression.
The extracted \ROO is found to be compatible with a recent measurement by the CMS collaboration~\cite{CMSROO} within the respective uncertainties.

The right side of Fig.~\ref{fig:ROOMeas} shows $R_\text{AA}$ at $p_\mathrm{T} = 9$\,GeV/$c$ for different collision systems as a function of $A^{1/3}$ as an approximation for the system size, with $A$ representing the mass number of the respective nucleus. 
A linear parametrization suggests an approximate scaling of the suppression with system size, which may reflect the dependence of energy loss on the in-medium path length. 
However, this interpretation is subject to several caveats: cold nuclear matter effects are not properly accounted for, collisional and radiative energy loss contributions are intertwined, and $A^{1/3}$ is only an approximate proxy for the path length.

\subsection{Comparisons of the nuclear modification factor to models with and without energy loss}

\begin{figure}
\begin{minipage}{0.49\linewidth}
\centerline{\includegraphics[width=\linewidth]{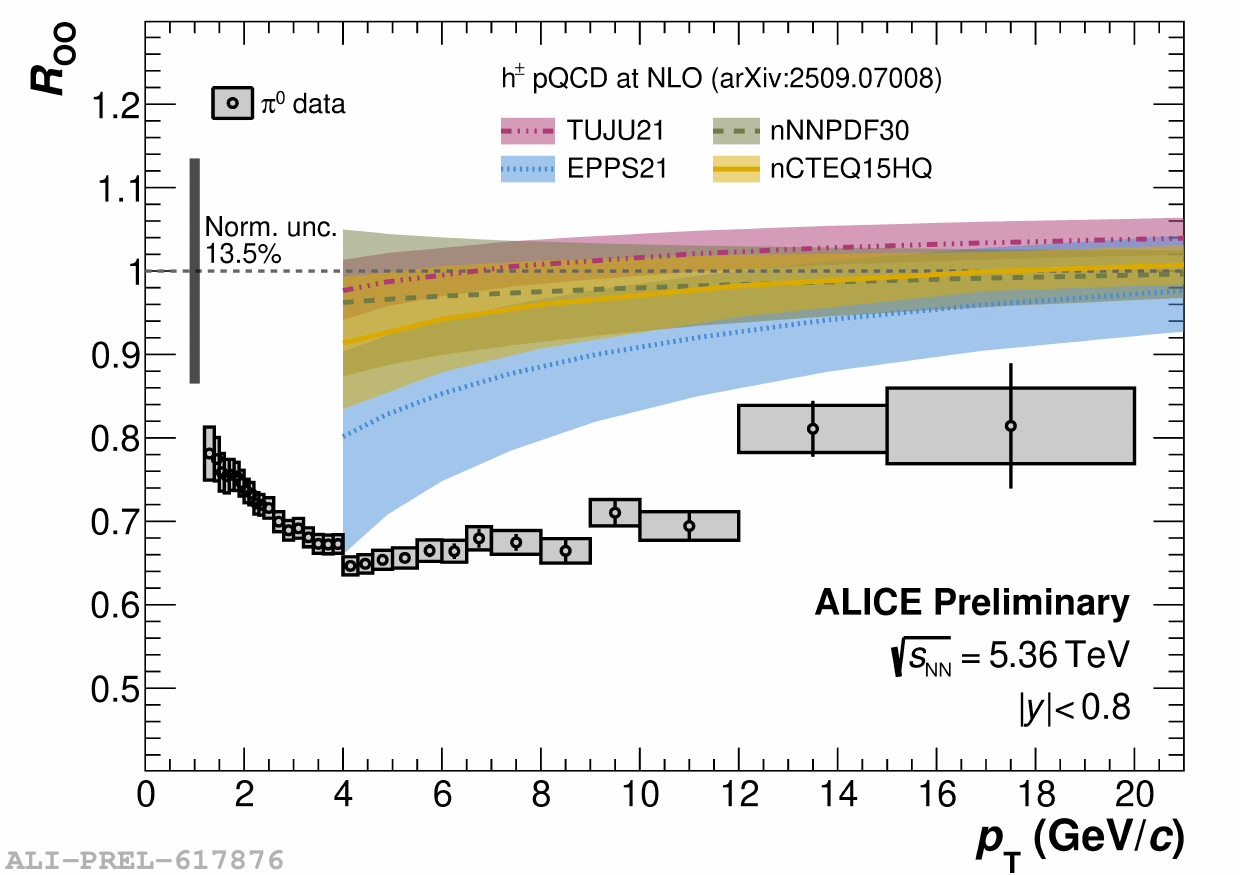}}
\end{minipage}
\hfill
\begin{minipage}{0.49\linewidth}
\centerline{\includegraphics[width=\linewidth]{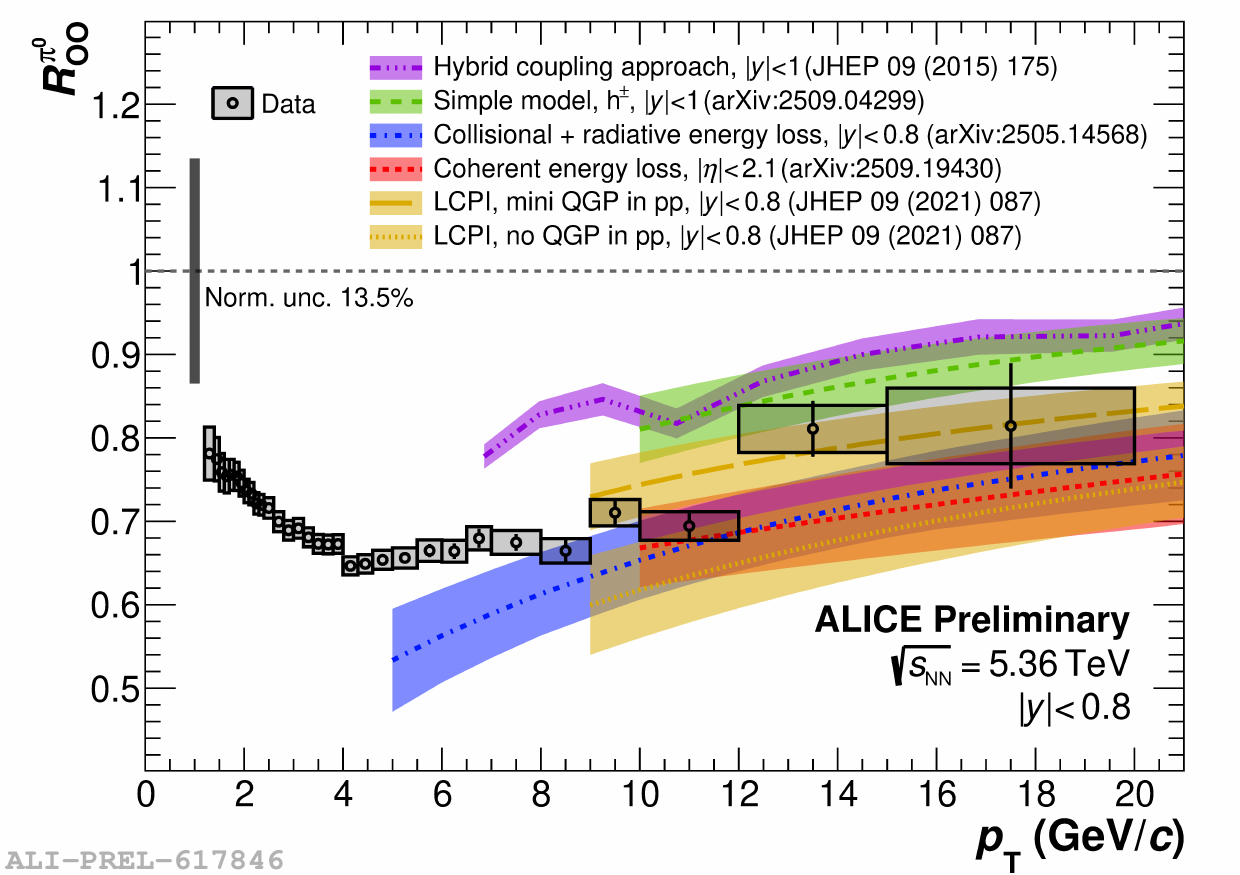}}
\end{minipage}
\caption[]{Comparison of measured \ROO to models without (left) and with (right) energy loss.}
\label{fig:ROOTheory}
\end{figure}

The left side of Fig.~\ref{fig:ROOTheory} compares the measured nuclear modification factor \ROO to pQCD predictions at NLO without energy loss~\cite{Mazeliauskas:2025clt} using four different nPDFs. Due to limited experimental constraints on the A-dependence of nPDFs, their predicted suppression varies significantly, ranging from a few percent (TUJU21) up to about 20\% (EPPS21) at low \pt, with sizeable theoretical uncertainties. As a result, the deviation of the data from these no-energy-loss baselines corresponds to a moderate significance of approximately $2.4\sigma$.

The right side of Fig.~\ref{fig:ROOTheory} shows comparisons to models including parton energy loss~\cite{ELossHybrid,ELossSimple,ELossCollRad,ELossCoherent,ELossLCPI}. While the models exhibit some spread, they all describe the data within the uncertainties. It should be noted that these models employ different treatments of cold nuclear matter effects, ranging from EPS09 at LO (LCPI) to EPPS21 at NLO (Hybrid).

\section{Outlook}

The large spread and uncertainties of current nPDFs limit the sensitivity of $R_\text{OO}$ to parton energy loss, as illustrated in Fig.~\ref{fig:ROOTheory}. 
To better constrain cold nuclear matter effects, the analysis of the recorded pO data is ongoing to extract \RpO.
Energy loss effects in \ROO can then be isolated using the double ratio \ROOpOpO, as suggested in a recent publication~\cite{nPDFPredictionsPaper}. This observable largely cancels cold nuclear matter contributions and provides a more direct probe of parton energy loss in OO collisions. 
A publication including \RpO and the double ratio is currently in preparation.


\section*{References}
\bibliography{moriond}

\end{document}